\documentclass[aps,pra,twocolumn,floatfix,superscriptaddress]{revtex4-2}
\usepackage[utf8]{inputenc}
\usepackage{mathrsfs}
\usepackage{graphicx}
\usepackage[english]{babel}
\usepackage{amsmath}
\usepackage{amssymb}
\usepackage{xcolor}
\usepackage{relsize}
\usepackage{textcomp}
\usepackage{dsfont}
\usepackage{bm}
\usepackage{float}
\renewcommand{\parallel}{\mathrel{/\mkern-4mu/}}

\newcommand{\me}{\mathrm{e}}      
\newcommand{\mi}{\mathrm{i}}      

\newcommand{\dif}{\mathrm{d}}     

\allowdisplaybreaks

\begin{document}

\title{
Torsion-induced gauge structure in curved quantum waveguides}
\author{Xu-Yang Hou}
\affiliation{School of Physics, Southeast University, Jiulonghu Campus, Nanjing 211189, China}
\author{Xianlong Gao}
\email{gaoxl@zjnu.edu.cn}
\affiliation{Department of Physics, Zhejiang Normal University, Jinhua 321004, China}
\author{Hao Guo}
\email{guohao.ph@seu.edu.cn}
\affiliation{School of Physics, Southeast University, Jiulonghu Campus, Nanjing 211189, China}
\affiliation{Hefei National Laboratory, University of Science and Technology of China, Hefei 230088, China}

\begin{abstract}
We investigate the effective dynamics of a particle confined near a
space curve. In the strict thin-layer reduction of the nondegenerate
transverse ground state, torsion does not enter the local effective
Hamiltonian, which contains only the curvature-induced scalar geometric
potential. In contrast, for a thin guide with finite transverse width,
a leading-order adiabatic projection onto the twofold-degenerate
first-excited transverse band renders the rotation of the Frenet normal
frame dynamically relevant and generates a matrix-valued Abelian gauge
potential.
Using a projection-based derivation in a co-rotating Frenet-frame basis, we show that this effective gauge potential is directly determined by the local torsion of the curve.
The resulting effective Hamiltonian takes a gauge-covariant form and produces two transverse-mode branches whose parabolic dispersions are shifted in opposite directions in momentum space.
For closed curves, the associated holonomy is controlled by the integrated torsion and leads to geometric interference.
These results provide a direct realization of a Wilczek--Zee-type connection induced purely by spatial geometry in curved quantum waveguides.
We further construct a classical-wave analogue using the degenerate bending modes of an isotropic elastic rod, demonstrating that the same torsion-induced gauge structure appears in continuum wave physics.
\end{abstract}

\maketitle

\section{Introduction}

The motion of a particle confined to a lower-dimensional submanifold of three-dimensional Euclidean space is a classic problem in both classical and quantum mechanics. In classical mechanics, two complementary approaches are commonly employed. The Newtonian method considers the particle as moving freely in the ambient space while being subjected to constraining forces that keep it on the desired surface or curve; these forces are typically taken to be normal to the constraint manifold, ensuring no work is done. The Lagrangian method eliminates the constraint forces from the outset by introducing generalized coordinates that parametrize the constrained manifold directly, leading to a reduced description solely in terms of the intrinsic geometry. For purely spatial constraints, both approaches yield equivalent equations of motion.

In quantum mechanics, however, the situation is far more subtle. The uncertainty principle prevents localizing the particle exactly on the constraint manifold; instead, a limiting process must be considered in which an increasingly strong transverse potential squeezes the wavefunction onto the submanifold. This limiting procedure is not guaranteed to be well-defined and may depend on the details of the confining potential. Early work by Jensen and Koppe \cite{Jensen1971} laid the foundation for understanding constrained quantum systems, and da Costa \cite{PhysRevA.23.1982,PhysRevA.25.2893} subsequently resolved the key technical issues by showing that, provided the transverse potential is chosen to be purely normal (depending only on the distance to the submanifold), the limit yields a unique effective Hamiltonian. For a particle constrained to a curve, da Costa found that the effective Hamiltonian contains a scalar geometric potential proportional to $-\kappa^2$, where $\kappa$ is the curvature, while the torsion $\tau$ does not appear in the local dynamics. This geometric potential depends on the extrinsic curvature of the embedded curve rather than on the intrinsic geometry alone. Consequently, curves that are intrinsically equivalent can exhibit different quantum mechanical properties, in sharp contrast to the corresponding classical constrained motion. This approach, now standard for curvature-induced quantum effects in low-dimensional systems, has been extended to particles in electromagnetic fields \cite{Ferrari2008}, to spin-1/2 particles on curved surfaces \cite{Wang2014}, and has been reviewed comprehensively in the literature \cite{Silva2015}.

Connection and gauge structures in reduced quantum descriptions have
long been recognized. In a different scattering-theoretical context,
Kuperin et al. studied connections associated with few-body quantum
scattering \cite{Kuperin1991}. Within constrained quantum mechanics,
geometry- and twist-induced gauge fields were developed in
Refs.~\cite{MaranerDestri1993,Mitchell2001}, while adiabatic
formulations identified Berry connections arising from the variation
of normal eigenfunctions
\cite{WachsmuthTeufel2010,WachsmuthTeufel2014}. In quantum waveguides,
transverse-mode elimination, mode mixing, and multimode dynamics have
likewise been formulated in terms of matrix-valued gauge connections,
including contributions from curvature, torsion, and longitudinal
variations of the transverse confinement
\cite{UlreichZwerger1998,StockhofeSchmelcher2014}. Curvature, torsion,
and cross-section twisting have also been studied extensively in the
mathematical theory of thin quantum waveguides
\cite{Bouchitte2007,ExnerKovarik2015}.

Torsion-dependent effects may therefore arise when transverse angular
momentum, mode degeneracy, or additional internal structure is
retained. Examples include global phase contributions in closed
twisted quantum rings \cite{Takagi1992} and geometry-induced gauge
effects in systems with spin--orbit coupling on curved surfaces and
space curves \cite{Magarill2003,Ortix2015}.

In the present work, we retain the twofold-degenerate first-excited
subspace of an isotropic two-dimensional harmonic confinement as the
transverse doublet. In this subspace, the rotation of the Frenet normal
frame induces a torsion-dependent Wilczek--Zee connection even in the
absence of spin--orbit coupling \cite{Wilczek1984}. Specifically, the
torsion $\tau(s)$ enters the projected dynamics through the
matrix-valued Abelian gauge potential
$\mathcal{A}_s=\hbar\tau(s)\sigma_y$, where $\sigma_y$ acts in the
degenerate transverse subspace. This connection is Abelian because it involves only a single generator, $[\mathcal{A}_s(s),\mathcal{A}_s(s')]=0$, yet it acts on a two-dimensional Hilbert space, so the holonomy is a $U(1)$ holonomy embedded in the corresponding $SU(2)$ representation. Its consequences include a torsion-controlled momentum shift of the two helicity branches, a correlation between propagation direction and helicity rather than strict helicity locking, and a geometric winding parameter that controls quantum interference.

Geometrically induced gauge potentials have been experimentally observed in photonic topological crystals \cite{Szameit2010} and in electronic states of curved nanostructures \cite{Onoe2012}. Non-Abelian $SU(2)$ gauge potentials have also been realized with ultracold atoms, enabling measurements of Wilson loops associated with geometric holonomies \cite{Sugawa2021}. These examples show that geometry-induced gauge structures are not merely formal theoretical constructions but can produce observable physical effects. Viewed in this broader context, the torsion-induced connection studied here provides a concrete example of how the embedding geometry of a curve can act as an effective gauge degree of freedom. We also construct a classical-wave analogue using the degenerate bending modes of an isotropic elastic rod, showing that the same torsion-induced gauge structure appears in continuum wave physics and may be accessible experimentally.

Our work is organized as follows. In Sec.~II we start
from the standard Frenet geometry of space curves and derive the
associated normal-bundle connection in the two-dimensional Frenet
normal frame, where torsion appears as the connection coefficient.
Section III presents the strict thin-layer reduction for the scalar
ground state and the leading-order adiabatic projection onto the
finite-width degenerate transverse band, culminating in the effective
Hamiltonian with the Wilczek--Zee connection. Section~IV
treats the helical quantum wire as a solvable model, computing the
energy spectrum, group velocity, and geometric phase. In Sec.~V we
discuss a classical-wave analogue using an isotropic elastic rod.
Finally, Sec.~VI concludes with a discussion of the geometric nature
of the torsion-induced gauge structure and its experimental
implications.

\section{Geometric framework and normal bundle connection}\label{Sec.2}

We begin with the standard Frenet--Serret description
of a smooth space curve and derive the associated normal-bundle
connection in the two-dimensional Frenet normal frame. The family of
normal planes along a space curve forms a rank-two normal bundle, whose
rotation is described by a natural connection.
In this representation, the torsion $\tau(s)$ appears
as the connection coefficient and provides the geometric input for the
subsequent projection onto a degenerate transverse subspace.

Consider a smooth curve $C$ in $\mathbb{R}^3$ parametrized by arc length $s$. The unit tangent vector is $\boldsymbol{T}(s)=\dif\boldsymbol{r}/\dif s\equiv\dot{\boldsymbol{r}}$. Curvature $\kappa(s)$ and principal normal $\boldsymbol{N}(s)$ are defined by $\dif\boldsymbol{T}/\dif s = \kappa(s)\boldsymbol{N}(s)$, and the binormal is $\boldsymbol{B}(s)=\boldsymbol{T}(s)\times\boldsymbol{N}(s)$. The Frenet-Serret equations in matrix form are
\begin{equation}\label{Frenet3}
\frac{\dif}{\dif s} \begin{bmatrix} \boldsymbol{T} \\ \boldsymbol{N} \\ \boldsymbol{B} \end{bmatrix} =
\begin{bmatrix}
0 & \kappa & 0 \\
-\kappa & 0 & \tau \\
0 & -\tau & 0
\end{bmatrix}
\begin{bmatrix} \boldsymbol{T} \\ \boldsymbol{N} \\ \boldsymbol{B} \end{bmatrix},
\end{equation}
and the torsion is $\tau(s)=(\dot{\boldsymbol{r}}\times\ddot{\boldsymbol{r}})\cdot\dddot{\boldsymbol{r}}/|\dot{\boldsymbol{r}}\times\ddot{\boldsymbol{r}}|^2$.

The normal plane at a point $s$ is spanned by $\boldsymbol{N}(s)$ and $\boldsymbol{B}(s)$. Collectively, these planes form the \textit{normal bundle} $NC$ over the curve.
The rotation of the Frenet normal frame induces a natural connection on $NC$.
As we shall see later, in a co-rotating transverse basis this geometric connection becomes a Wilczek–Zee gauge potential in the degenerate subspace. To extract this potential, we examine how a section $\boldsymbol{\xi}(s)=\xi_1(s)\boldsymbol{N}(s)+\xi_2(s)\boldsymbol{B}(s)$ of the normal bundle is transported along $s$. The ordinary derivative is
\begin{equation}
\frac{\dif\boldsymbol{\xi}}{\dif s} = \frac{\dif\xi_1}{\dif s}\boldsymbol{N} + \xi_1\frac{\dif\boldsymbol{N}}{\dif s} + \frac{\dif\xi_2}{\dif s}\boldsymbol{B} + \xi_2\frac{\dif\boldsymbol{B}}{\dif s}.
\end{equation}
Using the Frenet-Serret equations, $\dif\boldsymbol{N}/\dif s = -\kappa\boldsymbol{T}+\tau\boldsymbol{B}$ and $\dif\boldsymbol{B}/\dif s = -\tau\boldsymbol{N}$, we obtain
\begin{equation}
\frac{\dif\boldsymbol{\xi}}{\dif s} = \frac{\dif\xi_1}{\dif s}\boldsymbol{N} + \xi_1(-\kappa\boldsymbol{T}+\tau\boldsymbol{B}) + \frac{\dif\xi_2}{\dif s}\boldsymbol{B} - \xi_2\tau\boldsymbol{N}.
\end{equation}
The tangential component $-\kappa\xi_1\boldsymbol{T}$ lies outside the normal fibre.
The normal connection $\nabla^\perp_{\partial_s}$ projects the ordinary derivative onto the normal plane by discarding this tangential component:
\begin{equation}
\nabla^\perp_{\partial_s}\boldsymbol{\xi} = \left(\frac{\dif\xi_1}{\dif s} - \tau\xi_2\right)\boldsymbol{N} + \left(\frac{\dif\xi_2}{\dif s} + \tau\xi_1\right)\boldsymbol{B}.
\end{equation}
In matrix form with respect to the basis $(\boldsymbol{N},\boldsymbol{B})$,
\begin{equation}
\nabla^\perp_{\partial_s} \begin{pmatrix} \xi_1 \\ \xi_2 \end{pmatrix} = \left[\frac{\dif}{\dif s} + \begin{pmatrix} 0 & -\tau(s) \\ \tau(s) & 0 \end{pmatrix}\right]\begin{pmatrix} \xi_1 \\ \xi_2 \end{pmatrix}.
\end{equation}
The connection matrix is proportional to the Pauli matrix $\sigma_y$, because
\begin{equation}
\begin{pmatrix} 0 & -\tau \\ \tau & 0 \end{pmatrix} = -\mi\tau\sigma_y,\qquad \sigma_y = \begin{pmatrix} 0 & -\mi \\ \mi & 0 \end{pmatrix}.
\end{equation}
Hence the induced connection $1$-form on the normal bundle is
\begin{equation}
\omega = -\mi\tau(s)\sigma_y\,\dif s.
\end{equation}
Thus, in the Frenet normal-frame representation, the normal-bundle connection is an Abelian $U(1)$ connection embedded in the two-dimensional $SU(2)$ representation.
Because the base curve is one-dimensional, the curvature two-form vanishes identically:
\begin{equation}
\Omega = \dif\omega + \omega\wedge\omega = -\mi\frac{\dif\tau}{\dif s}\,\dif s\wedge\dif s + \omega\wedge\omega = 0,
\end{equation}
where $\dif s\wedge\dif s=0$ and $\omega\wedge\omega=0$ follows from the Abelian nature.
The normal connection is locally flat but can produce a nontrivial holonomy along a closed curve:
\begin{equation}
U_{\perp} = \mathcal{P}\exp\!\left(\oint\omega\right) = \exp\!\left[-\mi\sigma_y\oint \tau(s)\,\dif s\right].
\end{equation}
In the $\sigma_y$ eigenbasis, this holonomy reduces to opposite Abelian geometric phases determined by the integrated torsion.
When a degenerate transverse subspace is retained, this geometric connection appears as a Wilczek–Zee-type gauge potential in the projected quantum dynamics.

For a classical point particle constrained to the curve, the free Lagrangian $L=m\dot{s}^{2}/2$ contains neither $\kappa$ nor $\tau$.
The normal-bundle connection becomes dynamically relevant only when the system carries a normal-plane degree of freedom, such as a transverse orientation or a degenerate transverse mode.
In the quantum problem considered next, a twofold-degenerate transverse subspace provides such a fibre, allowing the geometric connection to appear as the gauge potential $\mathcal{A}_s = \hbar\tau(s)\sigma_y$.

\section{Thin-Waveguide Reduction and Effective Hamiltonians}\label{Sec.3}

\subsection{Scalar particle in the thin-layer limit}

Building on the normal-bundle geometry of the previous section, we now derive the effective dynamics of a particle confined to a space curve by thin-layer quantization. The procedure starts from the three-dimensional Schrödinger equation with a strong transverse potential and takes the limit of vanishing transverse width~\cite{PhysRevA.23.1982}. We first consider the scalar (nondegenerate) ground-state reduction. As derived in Appendix~\ref{scalar}, the effective Hamiltonian contains only a curvature-induced geometric potential, consistent with the known thin-layer result~\cite{PhysRevA.23.1982}. This provides the reference for the degenerate case in the next subsection.

Consider a particle of mass $m$ confined near a curve by a potential $V_{\perp}(u,v)$ that depends only on the transverse coordinates $(u,v)$ (the normal plane). The full Hamiltonian is
\begin{equation}
H = -\frac{\hbar^2}{2m}\nabla^2 + V_{\perp}(u,v).
\end{equation}
As discussed in Ref.~\cite{PhysRevA.23.1982}, a purely normal potential (constant on surfaces parallel to the curve) is essential for a well-defined limit; tangential dependence would produce spurious contributions. We introduce curvilinear coordinates adapted to the curve:
\begin{equation}
\boldsymbol{r}(s,u,v) = \boldsymbol{r}_0(s) + u\boldsymbol{N}(s) + v\boldsymbol{B}(s),
\end{equation}
where $\boldsymbol{r}_0(s)$ is the arc-length parametrization, and $(\boldsymbol{N},\boldsymbol{B})$ are the Frenet frame (principal normal and binormal). The coordinates $u$ and $v$ measure transverse displacements along $\boldsymbol{N}$ and $\boldsymbol{B}$, respectively. The potential is taken as a strong harmonic oscillator $V_{\perp}(u,v)=\frac{1}{2}m\omega^2(u^2+v^2)$; in the thin-layer limit $\ell_{\perp}=\sqrt{\hbar/m\omega}\to0$, the particle is forced to stay near the curve ($u,v\approx0$), freezing the transverse degrees of freedom and leaving only the longitudinal dynamics along $s$.

The validity of reduction relies on two small parameters: (i) the geometric thin-layer condition $\kappa\ell_\perp\ll1$, $\tau\ell_\perp\ll1$, which ensures that the transverse wavefunction is confined to a tubular neighbourhood where the curvilinear expansion is controlled; and (ii) the adiabatic condition $\hbar^2 k^2/(2m)\ll\hbar\omega$, which suppresses nonadiabatic transitions out of the selected transverse subspace. Both are satisfied in the limit $\omega\to\infty$ at fixed longitudinal momentum $k$.

The tangent vectors are computed using the Frenet–Serret equations:
\begin{align}
\partial_s\boldsymbol{r} &= \boldsymbol{T} + u\frac{\dif\boldsymbol{N}}{\dif s} + v\frac{\dif\boldsymbol{B}}{\dif s}
= (1-\kappa u)\boldsymbol{T} + \tau u\boldsymbol{B} - \tau v\boldsymbol{N},\notag\\
\partial_u\boldsymbol{r} &= \boldsymbol{N},\quad \partial_v\boldsymbol{r} = \boldsymbol{B}.
\end{align}
The metric tensor $g_{ij} = \partial_i\boldsymbol{r}\cdot\partial_j\boldsymbol{r}$ is
\begin{equation}
g_{ij} = \begin{pmatrix}
(1-\kappa u)^2 + \tau^2(u^2+v^2) & -\tau v & \tau u \\
-\tau v & 1 & 0 \\
\tau u & 0 & 1
\end{pmatrix}.
\end{equation}
To first order in $u,v$, we have $g_{ss} \approx (1-\kappa u)^2$, $g_{su} = -\tau v$, $g_{sv} = \tau u$, and the determinant is
\begin{equation}
\sqrt{g} = \sqrt{\det(g_{ij})} = 1-\kappa u + \mathcal{O}(u^2,v^2).
\end{equation}
For the scalar reduction, we project onto the nondegenerate transverse ground state $\chi_0(u,v)$, which may be chosen real and even in $u$ and $v$ for the isotropic harmonic confinement. Since this subspace is one-dimensional, it carries no nontrivial normal-frame index on which the normal-bundle connection can act. The wavefunction therefore separates as $\Psi(s,u,v)=\psi(s)\chi_0(u,v)$. The kinetic energy operator is the Laplace–Beltrami operator:
\begin{equation}
T = -\frac{\hbar^2}{2m}\frac{1}{\sqrt{g}}\partial_i\bigl(\sqrt{g}\,g^{ij}\partial_j\bigr).
\end{equation}
Integrating over the transverse coordinates (the detailed calculation is presented in Appendix~\ref{scalar}) yields the effective Hamiltonian
\begin{equation}\label{scalarH}
H_{\text{eff}}^{\text{(scalar)}} = -\frac{\hbar^2}{2m}\frac{\dif^2}{\dif s^2} - \frac{\hbar^2}{8m}\kappa^2(s).
\end{equation}
Notably, the torsion
$\tau(s)$ does not appear in this scalar effective Hamiltonian, in agreement with the known result for a nondegenerate transverse ground state~\cite{PhysRevA.23.1982}.

Thus, in the scalar ground-state reduction, torsion does not generate a gauge potential. The projected subspace is one-dimensional and carries no internal degree of freedom, so the normal-bundle connection has no matrix structure to act on. A torsion-induced gauge structure appears only when a nontrivial internal space is retained, such as a degenerate transverse subspace. This case is analyzed in the next subsection.

\subsection{Degenerate first-excited transverse subspace and induced
gauge potential (Wilczek--Zee connection)}

Keeping the same isotropic confinement, we retain the twofold-degenerate first-excited transverse subspace in a curved waveguide with small but finite transverse width $\ell_\perp$. We assume that $\ell_\perp$ is small compared with the local geometric length scales, namely, $\kappa\ell_\perp\ll1$ and $|\tau|\ell_\perp\ll1$. Together with the adiabatic separation of the transverse and longitudinal motions, these conditions justify a leading-order projection onto the selected transverse doublet. Within this projected subspace, the rotation of the Frenet normal frame is represented by a matrix-valued Abelian gauge potential determined by the torsion.

The eigenstates $\chi_{n_u,n_v}(u,v)=\psi_{n_u}(u)\psi_{n_v}(v)$ have energies $E_{n_u,n_v}=\hbar\omega(n_u+n_v+1)$.  The first excited subspace $\mathcal{H}_{\perp}^{(1)}=\{|1,0\rangle,|0,1\rangle\}$ is twofold degenerate with energy $E_1=2\hbar\omega$. The coordinate-space wavefunctions are
\begin{align}
\chi_{1,0}(u,v) &\propto u\,\exp\!\Bigl(-\frac{m\omega}{2\hbar}(u^2+v^2)\Bigr),\notag\\
\chi_{0,1}(u,v) &\propto v\,\exp\!\Bigl(-\frac{m\omega}{2\hbar}(u^2+v^2)\Bigr),
\end{align}
which have definite parity: $\chi_{1,0}$ is odd in $u$ and even in $v$, while $\chi_{0,1}$ is even in $u$ and odd in $v$. Geometrically, these states correspond to transverse vibrations along $\boldsymbol{N}(s)$ and $\boldsymbol{B}(s)$. Choosing the transverse basis to co-rotate with the Frenet frame gives the identification
\begin{equation}
\chi_1(s)\equiv |1,0\rangle_s\leftrightarrow\boldsymbol{N}(s),\qquad
\chi_2(s)\equiv |0,1\rangle_s\leftrightarrow\boldsymbol{B}(s).
\end{equation}
With this choice, the $s$-dependence of $\chi_a$ encodes the rotation of the normal frame, so the geometric connection appears in the degenerate subspace as a gauge potential determined by $\tau(s)$.

Expanding the total wavefunction in the degenerate subspace,
\begin{equation}
\Psi(s,u,v,t)=\sum_{a=1}^2\psi_a(s,t)\,\chi_a(u,v;s).
\end{equation}
The three-dimensional Hamiltonian is $H=-\frac{\hbar^2}{2m}\Delta+V_{\perp}$, where $\Delta$ is the Laplace--Beltrami operator in curvilinear coordinates (see Appendix~\ref{scalar}). Under the adiabatic approximation (the transverse energy gap is large compared to longitudinal kinetic and geometric couplings), the system is projected onto $\mathcal{H}_{\perp}^{(1)}$, and the effective Hamiltonian matrix elements are
\begin{equation}
(H_{\text{eff}})_{ab}=\langle\chi_a|H|\chi_b\rangle_{\perp},
\end{equation}
with $\langle\cdot\rangle_{\perp}$ denoting integration over $u,v$. We decompose $H=H_{\perp}+H_{\parallel}+H_{\text{coup}}$ (see Appendix~\ref{scalar}), where
\begin{align}
H_{\perp} &= -\frac{\hbar^2}{2m}(\partial_u^2+\partial_v^2)+V_{\perp},\notag\\
H_{\parallel} &= -\frac{\hbar^2}{2m}\frac{1}{1-\kappa u}\partial_s\!\Bigl(\frac{1}{1-\kappa u}\partial_s\Bigr),
\end{align}
and $H_{\text{coup}}$ contains the cross terms from the off-diagonal metric components $g^{su}$, $g^{sv}$.
Because $\chi_a$ are eigenstates of $H_{\perp}$ with $s$-independent eigenvalue $E_1$, we have $\langle\chi_a|H_{\perp}|\chi_b\rangle_{\perp}=E_1\delta_{ab}$. In the present finite-width description, $\omega$ is large but finite, so $E_1=2\hbar\omega$ remains finite. Within the projected transverse
subspace, this common energy offset can be absorbed into the zero of energy and is henceforth omitted.

For a thin waveguide with finite transverse width, the transverse wavefunctions are localized near $u=v=0$. Expanding $(1-\kappa u)^{-2}=1+2\kappa u+3\kappa^2u^2+\cdots$, the linear term vanishes by parity, while the quadratic term gives a relative
correction of order $\kappa^2\langle u^2\rangle_0\sim(\kappa\ell_\perp)^2$.
Since $\kappa\ell_\perp\ll1$, this finite-width correction is neglected, and $H_{\parallel}$ reduces to the free-particle form $-\hbar^2\partial_s^2/(2m)$. However, the derivative $\partial_s$ in $H_{\parallel}$ also acts on $\chi_a$, and this effect, together with $H_{\text{coup}}$, gives rise to the gauge connection.
To compute the coupling, we substitute the expansion into the Schr\"odinger equation and then project the result onto the degenerate subspace. For the longitudinal part we use
\begin{equation}
\Delta_{\parallel}(\psi_b\chi_b)=\frac{1}{f}\partial_s\!\Bigl(\frac{1}{f}(\partial_s\psi_b \chi_b+\psi_b\partial_s\chi_b)\Bigr),
\end{equation}
Here $f=1-\kappa u$. Since $\kappa\ell_\perp\ll1$, we set
$f^{-1}\simeq1$ to leading order, obtaining
\begin{equation}
\Delta_{\parallel}(\psi_b\chi_b)\approx \partial_s^2\psi_b\cdot\chi_b+2\partial_s\psi_b\cdot\partial_s\chi_b+\psi_b\partial_s^2\chi_b.
\end{equation}
Projecting onto $\chi_a$ yields
\begin{align}
\langle\chi_a|\Delta_{\parallel}|\chi_b\rangle_{\perp}
=\delta_{ab}\partial_s^2+2\langle\chi_a|\partial_s\chi_b\rangle_{\perp}\partial_s+\langle\chi_a|\partial_s^2\chi_b\rangle_{\perp}.
\end{align}
The cross term $\Delta_{\text{cross}}$ contributes similarly, and together they combine to give a gauge-covariant derivative.

We define the Wilczek--Zee connection on the degenerate transverse subspace by
\begin{equation}
(\mathcal{A}_s)_{ab}
=
\mi\hbar\langle\chi_a|\partial_s\chi_b\rangle_{\perp}.
\end{equation}
Although Wilczek–Zee connections are generally non-Abelian, here the connection is Abelian because only a single generator appears.
From the Frenet--Serret equations, the transverse basis rotates as
\begin{equation}\label{AppBFS}
\partial_s\begin{pmatrix}\chi_1\\\chi_2\end{pmatrix}=\tau(s)\begin{pmatrix}0&1\\-1&0\end{pmatrix}\begin{pmatrix}\chi_1\\\chi_2\end{pmatrix}
=\mi\tau(s)\sigma_y\begin{pmatrix}\chi_1\\\chi_2\end{pmatrix},
\end{equation}
so that $\langle\chi_a|\partial_s\chi_a\rangle_{\perp}=0$,
\begin{equation}
\langle\chi_1|\partial_s\chi_2\rangle_{\perp}=-\tau(s),\quad
\langle\chi_2|\partial_s\chi_1\rangle_{\perp}=\tau(s).
\end{equation}
Consequently,
\begin{equation}
\mathcal{A}_s=\hbar\tau(s)\begin{pmatrix}0&-\mi\\\mi&0\end{pmatrix}=\hbar\tau(s)\sigma_y.
\end{equation}
Thus $\mathcal{A}_s$ is Hermitian and proportional to $\sigma_y$ everywhere, so $[\mathcal{A}_s(s),\mathcal{A}_s(s')]=0$; it is an Abelian $U(1)$ connection embedded in the $SU(2)$ representation of the transverse doublet. Up to $\hbar$, $\mathcal{A}_s$ reproduces the normal-bundle connection $\omega=-\mi\tau\sigma_y\,\dif s$ from Sec.~\ref{Sec.2}, showing that torsion appears as a gauge potential through the same geometric connection.

The intraband part is encoded in the connection $\Gamma_{s,ab}=\langle\chi_a|\partial_s\chi_b\rangle_{\perp}$, which generates the linear coupling to $\partial_s$ and the $\mathcal{A}_s^2$ term in the covariant derivative.
The out-of-subspace contribution $\langle\partial_s\chi_a|(1-P)|\partial_s\chi_b\rangle_{\perp}$ is suppressed by the transverse excitation gap and the geometric smallness of the transverse width; it becomes significant only when the Frenet frame rotates very rapidly (large $\tau$) or when the level spacing $\hbar\omega$ is small. Under the adiabatic approximation and to leading order in the
transverse-width expansion, this term is neglected. With this separation, the projected longitudinal $\Delta_{\parallel}$ and cross terms $\Delta_{\text{cross}}$ combine into the gauge-covariant form
\begin{align}
-\frac{\hbar^2}{2m}\langle\chi_a|(\Delta_{\parallel}+\Delta_{\text{cross}})|\chi_b\rangle_{\perp}
&=\frac{1}{2m}\bigl[(-\mi\hbar\partial_s-\mathcal{A}_s)^2\bigr]_{ab}\notag\\
&\quad -\frac{\hbar^2}{8m}\kappa^2(s)\delta_{ab}.
\end{align}
The curvature potential $-\hbar^2\kappa^2/(8m)$ emerges from the metric-dependent Laplace--Beltrami operator after the wavefunction rescaling required by the curvilinear measure, exactly as in the scalar reduction
(see Appendix A); it is a scalar term proportional to the identity
and therefore does not lift the degeneracy.
A direct calculation, analogous to the scalar case but taking into account the odd parity of the transverse modes, shows that the cross term $\Delta_{\text{cross}}$ combines with the longitudinal part to produce the gauge-covariant derivative $(-\mi\hbar\partial_s-\mathcal{A}_s)^2$. The detailed derivation for the degenerate subspace is provided in Appendix~\ref{cross}.
Accordingly, under the adiabatic approximation and to leading order in
the transverse-width expansion, the two-component spinor
$\psi=(\psi_1,\psi_2)^T$ satisfies the effective Schrödinger equation
\begin{equation}
\mi\hbar\frac{\partial\psi}{\partial t}=\left[\frac{1}{2m}\bigl(-\mi\hbar\partial_s-\mathcal{A}_s\bigr)^2-\frac{\hbar^2}{8m}\kappa^2(s)\right]\psi,
\end{equation}
i.e., the effective Hamiltonian
\begin{equation}\label{Heff}
H_{\text{eff}}=\frac{1}{2m}\bigl(-\mi\hbar\partial_s-\mathcal{A}_s\bigr)^2-\frac{\hbar^2}{8m}\kappa^2(s).
\end{equation}
Expanding,
\begin{align}
H_{\rm eff}
&=
-\frac{\hbar^2}{2m}\partial_s^2
+
\frac{\mi\hbar^2}{m}\tau(s)\sigma_y\partial_s \notag \\
&+
\frac{\mi\hbar^2}{2m}\bigl(\partial_s\tau(s)\bigr)\sigma_y
+
\frac{\hbar^2\tau^2(s)}{2m}
-
\frac{\hbar^2\kappa^2(s)}{8m}.
\label{Heff_expanded}
\end{align}
For curves with constant torsion (e.g., the helix below), $\partial_s\tau=0$. Although $\mathcal{A}_s$ is matrix-valued, it remains proportional to $\sigma_y$, so the holonomy is Abelian. For the nondegenerate transverse ground state, similar thin-layer analyses yield an effective Hamiltonian containing only the curvature potential, with torsion contributing at most a scalar term through cross-section anisotropy~\cite{Krejcirik2012}. In the present degenerate-subspace setting, by contrast, torsion appears as a matrix-valued Abelian gauge field.

\begin{figure}[ht]
\centering
\includegraphics[width=\columnwidth]{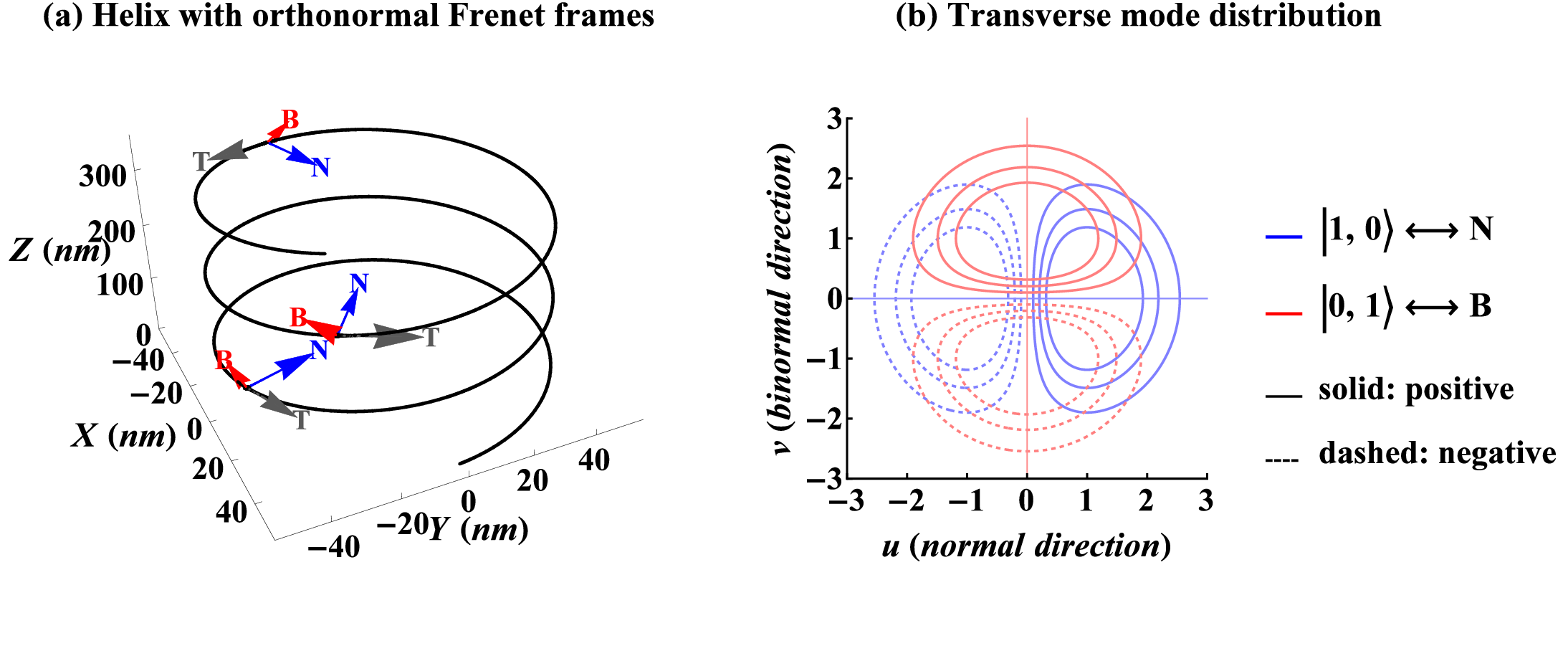}
\caption{(a) Helical curve with Frenet frames. The tangent $\boldsymbol{T}$,
principal normal $\boldsymbol{N}$, and binormal $\boldsymbol{B}$
are shown at representative points along the curve.
(b) Transverse mode distributions of the twofold-degenerate first
excited states, with $|1,0\rangle$ oriented along $\boldsymbol{N}$
and $|0,1\rangle$ oriented along $\boldsymbol{B}$. Solid and dashed
contours indicate positive and negative amplitudes, respectively. }
\label{Fig1}
\end{figure}

\section{Helical quantum wire: a solvable model}

As a concrete realization of the above general framework, we now consider a helical quantum wire. A helix is a space curve with constant curvature $\kappa$ and constant torsion $\tau$, making it an ideal testing ground for the torsion-induced gauge potential. The parametrization in terms of arc length $s$ is
\begin{equation}
\boldsymbol{r}(s) = \begin{pmatrix} R\cos(s/\rho) \\ R\sin(s/\rho) \\ \frac{P}{2\pi}\frac{s}{\rho} \end{pmatrix},\quad
\rho = \sqrt{R^2 + (P/2\pi)^2},
\end{equation}
where $R$ is the radius of the helix and $P$ its pitch. The curvature and torsion are constants given by
\begin{equation}
\kappa = \frac{R}{\rho^2},\qquad \tau = \frac{P/2\pi}{\rho^2}.
\end{equation}
Physically, $\kappa$ measures how tightly the curve bends, while $\tau$ measures how fast the osculating plane rotates around the tangent vector. As illustrated in Fig.~\ref{Fig1}(a), for a helix both quantities are constant; the uniform rotation of the Frenet frame at rate $\tau$ gives rise to a constant gauge potential $\mathcal{A}_s = \hbar\tau\sigma_y$. The spatial profiles of these two modes are illustrated in Fig.~\ref{Fig1}(b). The state $|1,0\rangle$ is odd in $u$ (along $\boldsymbol{N}$) and even in $v$, while $|0,1\rangle$ is odd in $v$ (along $\boldsymbol{B}$) and even in $u$. This geometric correspondence ensures that the transverse basis rotates with the Frenet frame, coupling the torsion to the internal degrees of freedom.

\begin{figure}[ht]
\centering
\includegraphics[width=\columnwidth]{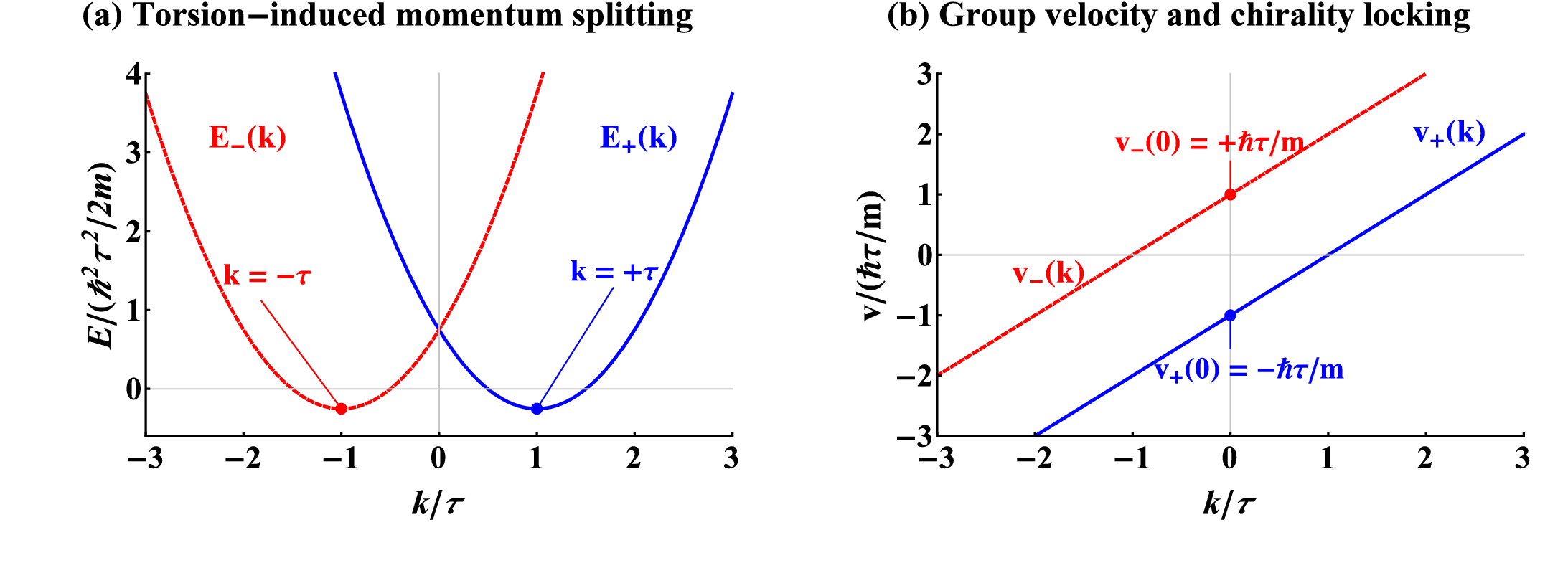}
\caption{
(a) Energy dispersion $E_\pm(k)$.
(b) Group velocity $\mathrm{v}_\pm(k)$.
In both panels, the blue solid curves represent the $+$ branch, whereas the red dashed curves represent the $-$ branch.
}
\label{Fig2}
\end{figure}

Substituting these constants into the effective Hamiltonian \eqref{Heff} yields
\begin{equation}
H_{\rm eff} = \frac{1}{2m}\bigl(-\mi\hbar\partial_s - \hbar\tau\sigma_y\bigr)^2 - \frac{\hbar^2}{8m}\kappa^2.
\end{equation}
Assuming plane-wave solutions $\psi(s) = \me^{\mi ks}\tilde{\psi}$ with $\tilde{\psi}$ a two-component spinor, we obtain
\begin{align}
H(k) =& \frac{\hbar^2}{2m}(k - \tau\sigma_y)^2 - \frac{\hbar^2}{8m}\kappa^2.
\end{align}
The eigenstates of $\sigma_y$ are $|\pm\rangle_y$ with eigenvalues $\pm1$, so the energy spectrum measured relative to the transverse energy $E_1$ is
\begin{equation}\label{spectrum}
E_\pm(k) = \frac{\hbar^2}{2m}(k \mp \tau)^2 - \frac{\hbar^2}{8m}\kappa^2.
\end{equation}
Thus the two helicity states have parabolic dispersions shifted by $\mp\tau$ in momentum space. The group velocity follows as
\begin{equation}
\mathrm{v}_\pm(k) = \frac{1}{\hbar}\frac{\partial E_\pm}{\partial k} = \frac{\hbar}{m}(k \mp \tau).
\end{equation}
At $k=0$, we have $\mathrm{v}_+ = -\hbar\tau/m$ and $\mathrm{v}_- = +\hbar\tau/m$. Thus, torsion induces a helicity-dependent momentum shift of the dispersion relations. As a result, near selected operating points (for example around $k=0$), the propagation direction becomes correlated with the helicity of the transverse mode ($|\pm\rangle_y$). This behavior reflects a helicity-dependent shift of the dispersion relations, which leads to a correlation between propagation direction and helicity near selected operating points rather than a globally locked transport mechanism.

As illustrated in Fig.~\ref{Fig2}(a), the torsion-induced momentum splitting appears as a horizontal shift of the two parabolic branches by $\mp\tau$. Figure~\ref{Fig2}(b) shows the corresponding group velocities $\mathrm{v}_\pm(k)$. At $k=0$, $\mathrm{v}_\pm=\mp\hbar\tau/m$, meaning that the two helicity branches have opposite group velocities at this selected momentum point. The zero crossings of $\mathrm{v}_\pm(k)$ occur at $k = \pm\tau$, respectively.

To estimate the magnitude of these effects for experimentally relevant parameters, consider a helical nanowire with radius $R=20\,\mathrm{nm}$ and pitch $P=125\,\mathrm{nm}$, for which $P/(2\pi)$ is comparable to $R$ and the torsion is appreciably enhanced. Then
\begin{align}
\rho &= \sqrt{20^2+(125/(2\pi))^2}\,\mathrm{nm} \approx 28.2\,\mathrm{nm},\notag\\
\tau &= \frac{125/(2\pi)}{\rho^2} \approx 2.5\times10^{-2}\,\mathrm{nm}^{-1}.
\end{align}
For an electron, the corresponding energy scale is $\hbar^2\tau^2/(2m_e)\approx2.4\times10^{-5}\,\mathrm{eV}$, equivalent to a temperature scale of about $0.28\,\mathrm{K}$. Although this remains a low-energy scale, the geometric phase accumulates rapidly over mesoscopic distances. For example, a $2\pi$ phase is reached over $L_{2\pi}=2\pi/\tau\approx0.25\,\mu\mathrm{m}$, and over $L=1\,\mu\mathrm{m}$ one obtains $\gamma=\tau L\approx25$. This suggests that the torsion-induced effect is more readily detected through phase-sensitive interference measurements, where the accumulated holonomy enters directly.

\begin{figure}[htbp]
\centering
\includegraphics[width=\columnwidth]{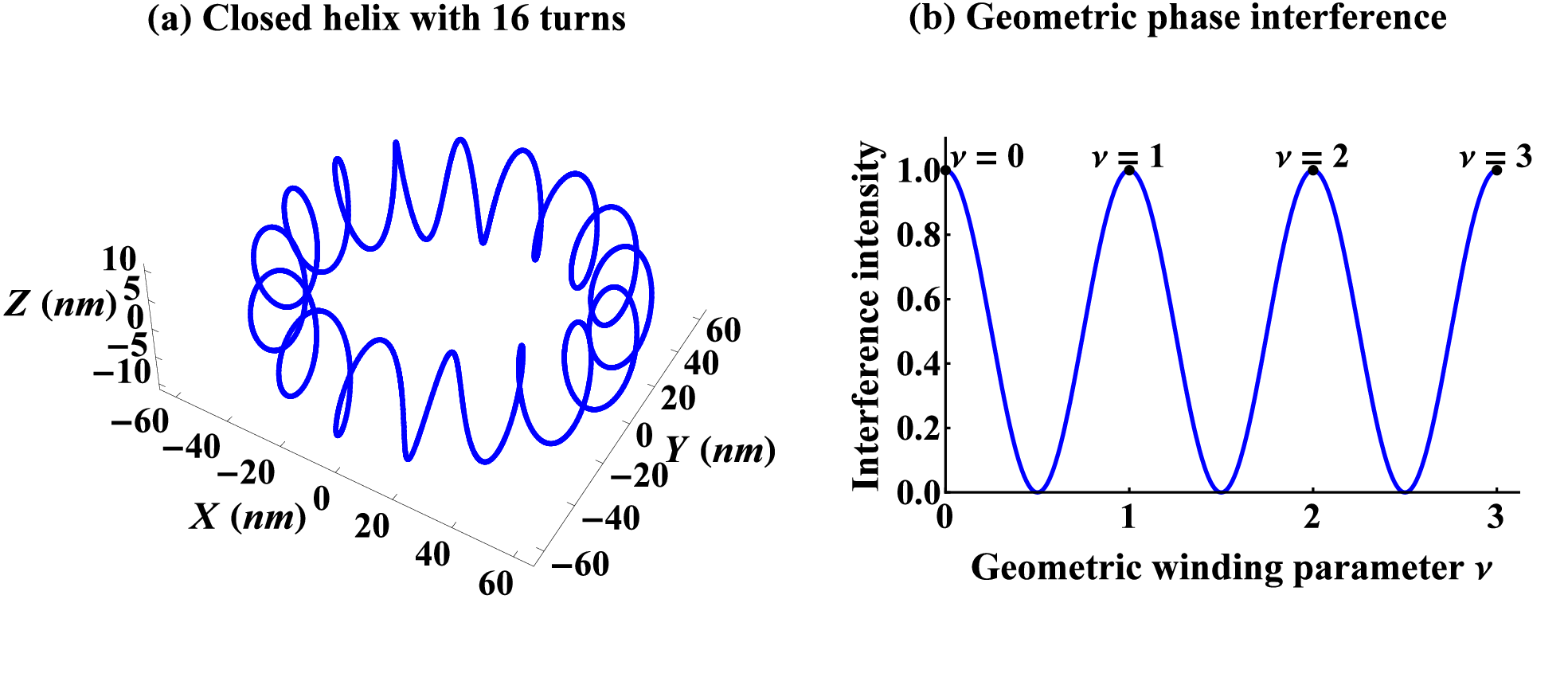}
\caption{(a) Closed helix with 16 turns. A particle traversing the loop acquires a geometric phase $\exp(\mi\tau L\sigma_y)$ due to the torsion-induced holonomy. (b) Interference intensity as a function of the geometric winding parameter $\nu = \tau L/(2\pi)$.}
\label{Fig3}
\end{figure}

We now examine the holonomy of the torsion-induced gauge potential, which plays a central role in interference phenomena. For a closed curve of length $L$ (with $s\in[0,L]$ and periodic boundary conditions), the evolution operator is
\begin{equation}
U(L) = \mathcal{P}\exp\left(\frac{\mi}{\hbar}\oint \mathcal{A}_s\dif s\right) = \exp\left(\mi\tau L\sigma_y\right),
\end{equation}
since $\mathcal{A}_s = \hbar\tau\sigma_y$ is constant and commutes with itself at different $s$. Expanding the exponential gives
\begin{equation}
U(L) = \cos(\tau L)\,\mathbb{I} + \mi\sin(\tau L)\,\sigma_y.
\end{equation}
Thus a state initially prepared in, say, $|+\rangle_y$ acquires a phase factor $\me^{\mi\tau L}$ after one round trip.

We define the geometric winding number
\begin{equation}
\nu = \frac{\tau L}{2\pi},
\end{equation}
which measures the total angle the Frenet frame rotates as one traverses the curve. A concrete example of a closed curve is shown in Fig.~\ref{Fig3}(a), where a helix is bent into a ring of 16 turns. After one full loop, the Frenet frame rotates by $\tau L = 2\pi\nu$, and the quantum state accumulates a holonomy $\exp(\mi\tau L\sigma_y)$. When $\nu$ is an integer, $U(L)=\mathbb{I}$, and the system returns to its initial state without any net geometric phase.

The physical consequence of this holonomy is a geometric interference pattern that depends on $\nu$. As shown in Fig.~\ref{Fig3}(b), if a particle is prepared in a superposition of the two helicity states and travels around a closed loop, the interference intensity oscillates with $\nu$: maximal constructive interference occurs at integer $\nu$, and destructive interference at half-integer $\nu$. This behaviour directly manifests the torsion-induced geometric phase. Note that $\nu$ is a continuously tunable parameter (by changing $L$ or the helix geometry), in contrast to quantized topological invariants such as the Chern number. This continuous tunability reflects the geometric, rather than topological, nature of the torsion-induced gauge structure.

\section{Classical-wave analogue of the geometric gauge structure}
\label{acoustic}

The geometric gauge structure derived from the Frenet frame is not restricted to quantum mechanics; it also admits classical-wave analogues. In this section we construct two minimal examples: a scalar acoustic mode in a curved
duct, which mirrors the scalar quantum particle on a space curve; and the degenerate bending modes of an isotropic elastic rod, which provide a natural classical-wave counterpart of the matrix-valued geometric connection. To expose the underlying covariant structure transparently, we adopt the following idealizations: circular cross-section, perfect isotropy, negligible dissipation, and the thin-layer (long-wavelength) limit where the transverse scale $d_\perp$ is much smaller than the curvature radius $\rho$ and the wavelength $\lambda$ is comparable to $\rho$.

\subsection{Scalar acoustic mode: curvature-induced geometric potential}

Consider a curved acoustic duct whose centerline follows a space curve $\boldsymbol{r}_0(s)$ with Frenet curvature $\kappa(s)$. In the thin-layer limit, the Helmholtz equation $\nabla^2 p = c^{-2}\partial_t^2 p$ reduces to an effective one-dimensional wave equation for the pressure amplitude $\phi(s,t)$ of the fundamental transverse mode, formally analogous to thin-layer quantization results for quantum particles constrained to curved manifolds~\cite{PhysRevA.23.1982,DuclosExner1995}:
\begin{equation}
\left(\partial_s^2+\frac{\kappa^2(s)}{4}\right)\phi = \frac{1}{c^2}\partial_t^2 \phi .
\label{scalar_acoustic}
\end{equation}
The transverse Laplacian contributes a constant cutoff frequency, while the longitudinal projection yields the geometric potential $-\kappa^2/4$. Notably, the torsion $\tau(s)$ does \emph{not} appear in Eq.~\eqref{scalar_acoustic}, because the scalar pressure field carries no internal polarization degree of freedom that could couple to the rotation of the Frenet frame. This absence is the exact classical counterpart of the quantum result for a scalar particle constrained to a space curve: torsion drops out of the effective dynamics, and only the curvature-induced geometric potential survives.

For time-harmonic modes $\phi(s,t)=\phi(s)\me^{-\mi\omega t}$, Eq.~\eqref{scalar_acoustic} becomes
\begin{equation}
\partial_s^2 \phi + \left(\frac{\omega^2}{c^2} + \frac{\kappa^2}{4}\right)\phi = 0,
\end{equation}
which maps directly onto the stationary Schr\"odinger equation for a scalar quantum particle with a curvature-induced potential, but with no gauge field.

\subsection{Degenerate bending modes: torsion as a matrix-valued geometric connection}

The non-trivial gauge structure emerges only when the wave field possesses a degenerate internal subspace, and the simplest classical-wave counterpart is an ideal isotropic elastic rod whose transverse displacement has two orthogonal polarizations. Let the rod centerline be $\boldsymbol{r}_0(s)$ with the Frenet frame $\{\boldsymbol{T},\boldsymbol{N},\boldsymbol{B}\}$ as defined in Sec.~\ref{Sec.2}. The curvature $\kappa(s)$ and torsion $\tau(s)$ satisfy the Frenet–Serret equations, which in component form read
\begin{equation}
\partial_s \boldsymbol{T} = \kappa\boldsymbol{N},\quad
\partial_s \boldsymbol{N} = -\kappa\boldsymbol{T} + \tau\boldsymbol{B},\quad
\partial_s \boldsymbol{B} = -\tau\boldsymbol{N},
\end{equation}
as already given in Eq.~(\ref{Frenet3}) of Sec.~\ref{Sec.2}. For curves where $\kappa(s)=0$ locally, a globally regular framing such as the Bishop frame (a rotation-minimizing frame that remains well-defined even when curvature vanishes)~\cite{Bishop1975} can be used.

The transverse displacement field $\boldsymbol{u}(s,t)$ lies in the normal plane:
\begin{equation}
\boldsymbol{u}(s,t) = \psi_1(s,t)\,\boldsymbol{N}(s) + \psi_2(s,t)\,\boldsymbol{B}(s) .
\label{displacement}
\end{equation}
For a circular cross-section with area $A$ and isotropic bending stiffness $EI$, the two bending directions are strictly degenerate in the straight-rod limit; we neglect warping, shear deformation, and material damping.
Applying $\partial_s$ to Eq.~\eqref{displacement} and using the Frenet–Serret relations, the longitudinal derivative of the displacement acquires a rotationally induced coupling in the transverse plane:
\begin{equation}
\bigl(\partial_s \boldsymbol{u}\bigr)_\perp = \bigl(\partial_s\psi_1 - \tau\psi_2\bigr)\boldsymbol{N} + \bigl(\partial_s\psi_2 + \tau\psi_1\bigr)\boldsymbol{B}.
\end{equation}
It is therefore natural to introduce the two-component amplitude $\boldsymbol{\psi}=(\psi_1,\psi_2)^T$ and the covariant derivative
\begin{equation}
D_s \boldsymbol{\psi} = \partial_s \boldsymbol{\psi} + \Gamma_s \boldsymbol{\psi},\quad
\Gamma_s = -\tau(s)\mi\sigma_y,
\end{equation}
where $\sigma_y$ is the Pauli matrix. This $\Gamma_s$ reproduces the quantum normal-bundle connection
$-\frac{\mi}{\hbar}\mathcal{A}_s$ appearing in the covariant derivative
$D_s=\partial_s-\frac{\mi}{\hbar}\mathcal{A}_s$ with $\mathcal{A}_s = \hbar\tau\sigma_y$. The connection $\Gamma_s$ is effectively Abelian because all $\Gamma_s$ are proportional to the same generator, so $[\Gamma_s,\Gamma_{s'}]=0$.

Within a minimal covariant effective description, the ordinary longitudinal derivative appearing in the Euler-Bernoulli bending energy is replaced by the covariant derivative induced by frame rotation: $\partial_s \to D_s = \partial_s + \Gamma_s$. Motivated by the covariant transport structure, we propose the following minimal covariant effective equation:
\begin{equation}
\rho A\,\partial_t^2 \boldsymbol{\psi} + EI\,D_s^4 \boldsymbol{\psi} = 0,
\label{EB_covariant}
\end{equation}
where $D_s^4$ denotes the fourth-order covariant derivative built from $D_s$. Equation~\eqref{EB_covariant} is a minimal effective model that captures the geometric transport structure; a complete elasticity theory of curved rods may contain additional curvature-induced terms, but they do not modify the covariant momentum shift generated by $\Gamma_s$.

To leading order in the adiabatic or WKB expansion (see Appendix~\ref{WKB}), we assume locally constant torsion and seek harmonic modes $\boldsymbol{\psi}(s,t)=\boldsymbol{\psi}_0\me^{\mi(k s-\omega t)}$. Diagonalising $\Gamma_s$ in the circular-polarization basis $\boldsymbol{e}_{\pm}=(\boldsymbol{N}\pm \mi\boldsymbol{B})/\sqrt{2}$ yields eigenvalues $\mp \mi\tau$, because $-\tau\mi\sigma_y$ has eigenvalues $\mp \mi\tau$ for $(\boldsymbol{N}\pm \mi\boldsymbol{B})/\sqrt{2}$. Hence $D_s$ acts as $\mi(k\mp\tau)$ in the respective eigenchannels. Substituting into Eq.~\eqref{EB_covariant} gives the dispersion relation
\begin{equation}
\omega_{\pm}(k) = \sqrt{\frac{EI}{\rho A}}\,(k\mp\tau)^2 .
\label{dispersion_acoustic}
\end{equation}
Equation~\eqref{dispersion_acoustic} provides a classical-wave counterpart of the quantum spectrum $E_{\pm}(k)=\frac{\hbar^2}{2m}(k\mp\tau)^2$. The momentum shift $k\to k\mp\tau$ induced by the geometric connection is the same in both systems. Their underlying dynamical equations are nevertheless different: in the elastic-wave model, the quadratic frequency dispersion arises from the fourth-order Euler--Bernoulli spatial operator~\cite{Graff1991}.

\subsection{Geometric phase and polarization transport}

A closely related geometric phase has been observed in helical acoustic waveguides, where sound vortices carrying orbital angular momentum acquire a spin-redirection geometric phase in momentum space~\cite{Wang2018SciAdv}.
More recently, an acoustic Pancharatnam--Berry phase arising from the polarization evolution of inhomogeneous acoustic velocity fields has been demonstrated experimentally using surface sound waves and acoustic metasurfaces~\cite{Xiao2026}. These studies provide experimental evidence that vector degrees of freedom in classical acoustic systems can support measurable geometric phases. The
mechanism considered here is nevertheless distinct: it originates from the real-space frame rotation of degenerate bending polarizations along a curved elastic rod. For a closed rod of total length $L$, the evolution operator for the polarization state is
\begin{equation}
U(L) = \mathcal{P}\exp\!\left(-\oint_0^L \Gamma_s\,\dif s\right)
= \exp\!\left( \mi\sigma_y\oint_0^L \tau(s)\,\dif s\right),
\end{equation}
since $-\Gamma_s = \tau\mi\sigma_y$ and the path ordering is trivial because all $\Gamma_s$ commute. An initially linearly polarized state (an equal superposition of the two bending components) emerges with a rotated polarization plane. The rotation angle is
\begin{equation}
\Delta\theta = \oint_0^L \tau(s)\,\dif s .
\label{polarization_rotation}
\end{equation}
This geometric rotation constitutes a classical geometric phase associated with frame transport. For a generic closed space curve the integral $\mathlarger{\oint}\tau(s)\,\dif s$ is a real number; it is not, in general, a topological invariant, even when the curve is equipped with a ribbon framing and the C\u{a}lug\u{a}reanu theorem (linking = twist + writhe)~\cite{DennisHannay2005} is invoked.

\subsection{Experimental feasibility}

Having established the theoretical correspondence, we now consider the conditions under which the torsion-induced gauge structure may be observed in a classical elastic rod. The most direct signature is the geometric rotation of the bending polarization described by Eq.~\eqref{polarization_rotation}.
For a circular helix of radius $R$ and reduced pitch $p=h/(2\pi)$, the curvature and torsion are constant, $\kappa=\frac{R}{R^{2}+p^{2}}$, $\tau=\frac{p}{R^{2}+p^{2}}$,
while the arc length accumulated over one turn is $s_{\rm turn}=2\pi\sqrt{R^{2}+p^{2}}$.
The geometric rotation acquired over one turn is therefore
\begin{equation}
\Delta\theta_{\rm turn} = \tau s_{\rm turn} =
\frac{2\pi p}{\sqrt{R^2+p^2}}
= 2\pi\cos\alpha,
\end{equation}
where $\alpha$ is the angle between the helix tangent and its axis, satisfying $\cos\alpha = p/\sqrt{R^2+p^2}$.
For a helix with $N$ turns,
\begin{equation}\label{helix_rotation}
\Delta\theta = 2\pi N\cos\alpha.
\end{equation}
The accumulated rotation is continuously tunable via
$R$, $p$, and $N$. By choosing these parameters so that
$2\pi N\cos\alpha$ is not close to an integer multiple of $\pi$, the output bending polarization may exhibit a distinguishable rotation relative to the reference polarization. The number of turns required in practice depends on the polarization resolution, attenuation, mode mixing, and boundary conditions of the experimental setup.

A key requirement is the preservation of the near-degeneracy of the two transverse bending modes. For a slightly elliptical cross section with ellipticity $\varepsilon$,
the two bending polarizations acquire a frequency splitting of relative magnitude $\frac{\delta\omega_\varepsilon}{\omega}
= O(\varepsilon)$.
By contrast, the torsion-induced momentum shift $k\rightarrow k\pm\tau$ produces a relative frequency splitting of order $\frac{\delta\omega_\tau}{\omega}
= O\!\left(\frac{\tau}{k_0}\right)$,
where $k_0$ denotes the operating wavenumber. Resolving the geometric splitting therefore requires
\begin{equation}
\varepsilon
\lesssim
O\!\left(\frac{\tau}{k_0}\right).
\label{isotropy_exp}
\end{equation}
This condition provides an order-of-magnitude criterion. Its quantitative implementation depends on the cross-sectional ellipticity, material anisotropy, elastic constants, operating frequency, and boundary conditions of the specific experimental system.

Material damping provides a second practical constraint. For an isotropic rod obeying Euler--Bernoulli dynamics, the dispersion relation is $\omega(k) = \sqrt{EI/(\rho A)}\,k^{2}$, so the group velocity becomes $v_g = 2\sqrt{EI/(\rho A)}\,k$. Using typical values for steel, $E\sim2\times10^{11}\,{\rm Pa}$, $\rho\sim8\times10^3\,{\rm kg/m^3}$, and a rod with diameter $d\sim1\,{\rm mm}$ and length $L\sim0.5\,{\rm m}$, the lowest bending mode corresponds to $k\sim\pi/L\approx6\,{\rm m^{-1}}$, yielding
\begin{equation}
v_g \sim 10\,{\rm m/s}.
\end{equation}
For metallic rods with $Q\sim10^{2}-10^{3}$, the damping length is expected to exceed the sample size by one to several orders of magnitude.
Consequently, attenuation is not expected to constitute the primary limitation for observing the geometric polarization transport.

A two-arm interferometer provides a simple geometry: a straight rod and a helix of equal length. A linearly polarized bending pulse launched into both arms yields a relative geometric rotation \eqref{helix_rotation} in the helical arm, while the straight arm retains its polarization. At the output of each arm, the two orthogonal transverse displacement components may be measured, for example by multi-component laser vibrometry, to reconstruct the bending-polarization angle. The difference between the reconstructed output angles then gives the relative geometric rotation.
In practice, ordinary dynamical phases and boundary-induced mode mixing can be reduced or calibrated using equal path lengths and identical clamping conditions. Because reversing the helix handedness changes $\tau\to-\tau$, the handedness-odd part of the measured rotation provides an additional means of isolating the
torsion-induced contribution.

Detecting the momentum-space shift $k\to k\pm\tau$ requires spectral resolution of the two circular-polarization branches and is more demanding. In contrast, the geometric rotation angle \eqref{helix_rotation} accumulates linearly with the number of turns, offering a more accessible experimental signature of the torsion-induced gauge structure.

\begin{table*}[ht]
\caption{Quantum-classical correspondence for geometric gauge transport on a space curve.}
\label{tab:correspondence}
\begin{ruledtabular}
\begin{tabular}{lcc}
Quantity & Quantum  & Classical analogue \\
\hline
Field & $\psi(s)$ & Transverse displacement $\boldsymbol{u}(s)$ \\
Mass / inertia & $m$ & Line density $\rho A$ \\
Geometric potential & $-\frac{\hbar^2}{8m}\kappa^2$ & $-\frac{1}{4}\kappa^2$ (scalar) \\
Geometric connection & $\mathcal{A}_s=\hbar\tau\sigma_y$ & $\Gamma_s=-\tau\mi\sigma_y$ \\
Covariant derivative & $D_s=\partial_s-\frac{\mi}{\hbar}\mathcal{A}_s$ & $D_s=\partial_s+\Gamma_s$ \\
Dispersion (shifted) & $E_{\pm}\propto (k\mp\tau)^2$ & $\omega_{\pm}\propto (k\mp\tau)^2$ \\
Polarization transport & Quantum phase $\gamma$ & Rotation angle $\Delta\theta$ \\
\end{tabular}
\end{ruledtabular}
\end{table*}

\subsection{Correspondence and distinctions}

Table~\ref{tab:correspondence} summarizes the quantum-classical correspondence established in this section. The structural correspondence resides in the covariant derivative $D_s=\partial_s+\Gamma_s$ and in the resulting
momentum-space shift.
The differences are equally important: the classical elastic system obeys a second-order time derivative, while the Euler--Bernoulli bending dynamics involves a fourth-order spatial operator; moreover, the strict degeneracy of the bending modes is an idealization that requires carefully engineered isotropy.

Conceptually related classical-wave realizations of geometric transport have been demonstrated in helical acoustic waveguides, where sound vortices carrying orbital angular momentum acquire momentum-space Berry phases induced by the macroscopic rotation of the waveguide~\cite{Wang2018SciAdv}. In that setting, the orbital angular momentum serves as an internal degree of freedom that couples to the helical geometry, analogous to the role played here by the degenerate bending polarizations. The key distinction is that the helical-waveguide Berry connection lives in momentum space and gives rise to a scalar $U(1)$ phase, whereas the Frenet-frame connection studied here operates in real space and produces a matrix-valued $SO(2)$ polarization rotation. Both constructions illustrate the universal principle that geometry induces effective gauge fields for vector waves, yet they differ in their parameter spaces and in the microscopic origin of the gauge structure.

From a broader perspective, the appearance of the covariant derivative $D_s=\partial_s+\Gamma_s$ reflects a universal geometric mechanism: whenever a degenerate vector wave is transported along a curved trajectory, the rotation of the local frame induces an effective gauge connection acting on the internal polarization space. The ideal isotropic elastic rod provides the minimal classical-wave analogue of this mechanism within our geometric framework. It captures the same matrix-valued gauge structure generated by torsion, yields a non-trivial momentum-space shift in the dispersion relation, and predicts an accumulated geometric rotation of the bending polarization. At the same time, the differences, particularly the fourth-order spatial operator and the classical wave dynamics, underscore that this is a structural analogue sharing the covariant geometry, rather than a strict physical isomorphism.

\section{Conclusion}

We have analyzed the distinct roles of curvature and torsion for a
quantum particle confined near a space curve. The nondegenerate
ground-state sector is treated by a strict thin-layer reduction,
whereas the first-excited transverse doublet at finite width is
described adiabatically to leading order. Within the general framework
of geometry-induced gauge connections, this degenerate-sector
construction provides an explicit two-dimensional Frenet-frame
realization of the normal-bundle connection. Curvature produces the scalar geometric potential $-\hbar^2\kappa^2/(8m)$, whereas torsion becomes dynamically active only when a degenerate transverse subspace is retained. In this case, the normal-bundle connection is represented in the transverse doublet as the matrix-valued Abelian Wilczek--Zee gauge potential $\mathcal{A}_s=\hbar\tau\sigma_y$. This potential does not open a gap but produces a torsion-controlled momentum displacement of the two helicity branches, thereby generating a direction--helicity correlation in propagation. For closed paths, the holonomy is governed by the integrated torsion, equivalently by the geometric winding number $\nu=\tau L/(2\pi)$.

These results clarify the geometric character of the effect. The torsion-induced gauge structure is controlled by the embedding geometry of the curve and the rotation of the Frenet normal frame, not by spin--orbit coupling or a quantized topological invariant. Thus, torsion provides a purely geometric route to helicity-resolved gauge effects.
We have further constructed a structural classical-wave analogue
based on the degenerate bending modes of an isotropic elastic rod, whose geometric description involves the same normal-bundle connection and associated covariant derivative. In a helical rod, the accumulated rotation of the bending polarization may provide an experimental signature of the torsion-induced gauge structure.

\begin{acknowledgments}
H. G. was supported by the Quantum Science and Technology-National
Science and Technology Major Project (Grant No. 2021ZD0301904) and the National Natural Science
Foundation of China (Grant No. 12447216). X.-Y. H. was supported by the National Natural Science
Foundation of China (Grant No. 12405008).
\end{acknowledgments}

\appendix

\section{Thin-layer quantization for a scalar particle}\label{scalar}

Here we  provide a rigorous derivation of the one-dimensional scalar effective Hamiltonian
\begin{equation}
H_{\text{eff}}^{\text{(scalar)}} = -\frac{\hbar^2}{2m}\frac{\dif^2}{\dif s^2} - \frac{\hbar^2}{8m}\kappa^2(s)
\end{equation}
following da Costa's thin-layer quantization method \cite{PhysRevA.23.1982}. All geometric quantities (curvature $\kappa$, torsion $\tau$) are retained to show why torsion does not appear in the final result.

We first introduce curvilinear coordinates adapted to the curve. Let the space curve $C$ be parametrized by arc length $s$ as $\boldsymbol{r}_0(s)$, with Frenet frame $\{\boldsymbol{T}(s),\boldsymbol{N}(s),\boldsymbol{B}(s)\}$. The coordinates near the curve are
\begin{equation}
\boldsymbol{r}(s,u,v) = \boldsymbol{r}_0(s) + u\boldsymbol{N}(s) + v\boldsymbol{B}(s),
\end{equation}
where $u$ and $v$ are Cartesian coordinates in the normal plane. The tangent vectors are
\begin{align}
\boldsymbol{e}_s &\equiv \partial_s\boldsymbol{r} = (1-\kappa u)\boldsymbol{T} - \tau v\boldsymbol{N} + \tau u\boldsymbol{B},\notag\\
\boldsymbol{e}_u &\equiv \partial_u\boldsymbol{r} = \boldsymbol{N},\quad
\boldsymbol{e}_v \equiv \partial_v\boldsymbol{r} = \boldsymbol{B}.
\end{align}
The metric components $g_{ij}=\boldsymbol{e}_i\cdot\boldsymbol{e}_j$ are
\begin{equation}
g_{ij} = \begin{pmatrix}
(1-\kappa u)^2 + \tau^2(u^2+v^2) & -\tau v & \tau u \\
-\tau v & 1 & 0 \\
\tau u & 0 & 1
\end{pmatrix},
\end{equation}
and $\det(g_{ij}) = (1-\kappa u)^2$, so $\sqrt{g}=|1-\kappa u|$. In the thin-layer limit $|\kappa u|\ll1$ we take $\sqrt{g}=1-\kappa u$. The inverse metric is
\begin{equation}\label{invg}
g^{ij} = \begin{pmatrix}
\dfrac{1}{(1-\kappa u)^2} & \dfrac{\tau v}{(1-\kappa u)^2} & -\dfrac{\tau u}{(1-\kappa u)^2} \\[8pt]
\dfrac{\tau v}{(1-\kappa u)^2} & 1 + \dfrac{\tau^2 v^2}{(1-\kappa u)^2} & -\dfrac{\tau^2 uv}{(1-\kappa u)^2} \\[8pt]
-\dfrac{\tau u}{(1-\kappa u)^2} & -\dfrac{\tau^2 uv}{(1-\kappa u)^2} & 1 + \dfrac{\tau^2 u^2}{(1-\kappa u)^2}
\end{pmatrix}.
\end{equation}
The explicit form of the Laplace--Beltrami operator can be expressed based on this. The three-dimensional kinetic operator is $T = -\frac{\hbar^2}{2m}\Delta$ with
\begin{equation}
\Delta = \frac{1}{\sqrt{g}}\partial_i\bigl(\sqrt{g}\,g^{ij}\partial_j\bigr).
\end{equation}
Substituting $\sqrt{g}=1-\kappa u$ and the inverse metric, $\Delta$ splits into longitudinal, transverse, and cross parts: $\Delta = \Delta_{ss} + \Delta_{uu} + \Delta_{vv} + \Delta_{\text{cross}}$, where
\begin{align}\label{Ddecom}
\Delta_{ss} &= \frac{1}{1-\kappa u}\partial_s\!\Bigl(\frac{1}{1-\kappa u}\partial_s\Bigr),\notag\\
\Delta_{uu} &= \frac{1}{1-\kappa u}\partial_u\!\Bigl[(1-\kappa u)\Bigl(1+\frac{\tau^2 v^2}{(1-\kappa u)^2}\Bigr)\partial_u\Bigr],\notag\\
\Delta_{vv} &= \frac{1}{1-\kappa u}\partial_v\!\Bigl[(1-\kappa u)\Bigl(1+\frac{\tau^2 u^2}{(1-\kappa u)^2}\Bigr)\partial_v\Bigr],\notag\\
\Delta_{\text{cross}} &= \frac{1}{1-\kappa u}\Big[
\partial_s\!\Bigl(\frac{\tau v}{1-\kappa u}\partial_u\Bigr)
+ \partial_u\!\Bigl(\frac{\tau v}{1-\kappa u}\partial_s\Bigr)\notag\\
&\qquad+ \partial_s\!\Bigl(-\frac{\tau u}{1-\kappa u}\partial_v\Bigr)
+ \partial_v\!\Bigl(-\frac{\tau u}{1-\kappa u}\partial_s\Bigr)
\Bigr].
\end{align}
The transverse cross terms involving $g^{uv}$ (proportional to $\tau^2 uv$) are of higher order in the thin-layer expansion and are neglected here. All expressions are valid to leading order in the small parameters $\kappa\ell_\perp$, $\tau\ell_\perp$.
To obtain a Hermitian one-dimensional operator after reduction, da Costa introduced the renormalized wavefunction
\begin{equation}
\Phi(s,u,v) = (1-\kappa u)^{1/2}\,\Psi(s,u,v).
\end{equation}
Setting $f=1-\kappa u$, we have $\Psi = f^{-1/2}\Phi$. A direct calculation yields the effective Laplacian $\tilde{\Delta}$ acting on $\Phi$,
\begin{align}\label{tildeDelta}
\tilde{\Delta}\Phi \equiv &f^{1/2}\Delta(f^{-1/2}\Phi) \notag\\
=& \frac{1}{f}\partial_s\!\Bigl(\frac{1}{f}\partial_s\Phi\Bigr) + \partial_u^2\Phi + \partial_v^2\Phi + \frac{\kappa^2}{4f^2}\Phi + \Delta_{\text{cross}}^{(1)}.
\end{align}
The term $\frac{\kappa^2}{4f^2}$ arises from the combination $-\frac{1}{2}\frac{\partial_i f}{f}g^{ij}\partial_j$ (which vanishes upon transverse average due to odd parity) and $+\frac{1}{4}\frac{\partial_i f\partial_j f}{f^2}g^{ij}$ (which gives $\frac{\kappa^2}{4f^2}$ after using $g^{ss}=1/f^2$ and $\partial_s f = -\kappa' u$). Here $\Delta_{\text{cross}}^{(1)}$ contains the renormalized cross terms, which involve $\tau$ and odd powers of $u,v$. These vanish upon averaging over the transverse ground state because the latter is even.

We now take the thin-layer limit and perform the transverse average. The transverse confining potential $V_{\perp}(u,v)$ (e.g., $V_{\perp}=\frac{1}{2}m\omega^2(u^2+v^2)$) forces the particle to stay near $u=v=0$. The transverse ground state $\chi_0(u,v)$ satisfies
\begin{equation}
\Bigl(-\frac{\hbar^2}{2m}(\partial_u^2+\partial_v^2)+V_{\perp}\Bigr)\chi_0 = E_0\chi_0,
\end{equation}
is real and even, and is normalized to unity. In the strong-confinement limit ($\omega\to\infty$), $\chi_0$ is highly localized. We set $\Phi(s,u,v,t)=\phi(s,t)\chi_0(u,v)$ and evaluate $\tilde{\Delta}\Phi$ at $u=v=0$ (where $f\to1$, $\kappa^2/(4f^2)\to\kappa^2/4$):
\begin{equation}
\tilde{\Delta}\Phi\big|_{u=v=0} = \partial_s^2\phi\cdot\chi_0 + \phi\cdot(\partial_u^2+\partial_v^2)\chi_0 + \frac{\kappa^2}{4}\phi\chi_0.
\end{equation}
Inserting this into the Schr\"odinger equation $-\frac{\hbar^2}{2m}\tilde{\Delta}\Phi = \mi\hbar\partial_t\Phi$, multiplying by $\chi_0^*$ and integrating over $u,v$, we obtain
\begin{equation}
-\frac{\hbar^2}{2m}\frac{\dif^2\phi}{\dif s^2} + E_0\phi - \frac{\hbar^2\kappa^2}{8m}\phi = \mi\hbar\frac{\partial\phi}{\partial t}.
\end{equation}
Dropping the constant $E_0$ (which can be absorbed into the energy zero) gives the one-dimensional effective equation
\begin{equation}
-\frac{\hbar^2}{2m}\frac{\dif^2}{\dif s^2}\phi(s,t) - \frac{\hbar^2}{8m}\kappa^2(s)\,\phi(s,t) = \mi\hbar\frac{\partial\phi}{\partial t},
\end{equation}
and hence the effective Hamiltonian
\begin{equation}
H_{\text{eff}}^{\text{(scalar)}} = -\frac{\hbar^2}{2m}\frac{\dif^2}{\dif s^2} - \frac{\hbar^2}{8m}\kappa^2(s).
\end{equation}

Finally, we explain why torsion does not appear. All torsion-dependent terms in Eq.~\eqref{tildeDelta} are contained in $\Delta_{\text{cross}}^{(1)}$, which has structures like $\partial_s(\tau v\partial_u)$ etc. When acting on $\Phi=\phi(s)\chi_0(u,v)$ and averaged transversely, they yield integrals such as $\mathlarger{\int} \chi_0 v\partial_u\chi_0\,\dif u\dif v$. Because $\chi_0$ is even in $u$ and $v$, $\partial_u\chi_0$ is odd in $u$, making the integrand odd in $u$; the integral vanishes. The same holds for any term containing an odd power of $u$ or $v$. Hence torsion does not contribute to the scalar effective Hamiltonian.
This is consistent with da Costa's result: a nondegenerate scalar particle experiences only the curvature-induced geometric potential. Within the scalar transverse ground-state thin-layer reduction considered here, torsion contributes neither a local gauge potential nor a global holonomy at the level of the effective one-dimensional dynamics.

\section{Projection onto the degenerate transverse subspace and emergence of the covariant derivative}\label{cross}

In this appendix, we derive the leading-order effective Hamiltonian in the twofold-degenerate first-excited transverse subspace using a direct projection method. The waveguide is taken to have a small but finite transverse width, with $\kappa\ell_\perp\ll1$ and $|\tau|\ell_\perp\ll1$, while the transverse level spacing
$\hbar\omega$ is assumed to be much larger than the longitudinal kinetic and geometric energy scales.

We employ the Frenet coordinates $(s,u,v)$ and the metric structure established in Appendix~\ref{scalar}. To leading order in the thin-waveguide expansion, the inverse metric components are given in Eq.~(\ref{invg}),
and the Laplace--Beltrami operator decomposes as $\Delta=\Delta_{\parallel}+\Delta_{\perp}+\Delta_{\text{cross}}$, with
\begin{align}
\Delta_{\parallel}=\frac{1}{f}\partial_s\!\Bigl(\frac{1}{f}\partial_s\Bigr),\quad
\Delta_{\perp}=\partial_u^2+\partial_v^2-\frac{\kappa}{f}\partial_u,
\end{align}
and $\Delta_{\text{cross}}$ is given in Eq.~(\ref{Ddecom})
The total wavefunction is expanded in the degenerate first-excited transverse subspace $\mathcal{H}_{\perp}^{(1)}=\{|1,0\rangle,|0,1\rangle\}$ as
\begin{equation}
\Psi(s,u,v)=\sum_{b=1}^{2}\psi_b(s)\chi_b(u,v;s).
\end{equation}
The basis states co-rotate with the Frenet frame:
\begin{equation}
\chi_1(s)\equiv|1,0\rangle_s\leftrightarrow\boldsymbol{N}(s),\quad
\chi_2(s)\equiv|0,1\rangle_s\leftrightarrow\boldsymbol{B}(s),
\end{equation}
with coordinate-space representations
\begin{equation}
\chi_1=Cu\me^{-\frac{\alpha}{2}(u^2+v^2)},\quad
\chi_2=Cv\me^{-\frac{\alpha}{2}(u^2+v^2)},
\end{equation}
where $\alpha=m\omega/\hbar$ and $C=\sqrt{\frac{2\alpha^2}{\pi}}$.
The co-rotating basis satisfies $\langle\chi_a|\chi_b\rangle_{\perp}=\delta_{ab}$ and evolves according to the Frenet--Serret equations (\ref{AppBFS}).
The associated Wilczek--Zee connection is defined by
\begin{equation}
(\Gamma_s)_{ab}=\langle\chi_a|\partial_s\chi_b\rangle_{\perp},\quad
\mathcal{A}_s=\mi\hbar\Gamma_s=\hbar\tau\sigma_y.
\end{equation}

The geometric gauge structure emerges directly from the projection of the longitudinal derivative onto the co-rotating basis. Acting with $\partial_s$ on $\Psi$ gives
\begin{equation}
\partial_s\Psi=\sum_b\bigl[(\partial_s\psi_b)\chi_b+\psi_b\partial_s\chi_b\bigr].
\end{equation}
Projecting onto $\chi_a$ yields
\begin{equation}
\langle\chi_a|\partial_s\Psi\rangle_{\perp}=\partial_s\psi_a+\sum_b(\Gamma_s)_{ab}\psi_b,
\end{equation}
so that the ordinary derivative is promoted to the covariant derivative
\begin{equation}
D_s=\partial_s+\Gamma_s=\partial_s-\frac{\mi}{\hbar}\mathcal{A}_s.
\end{equation}
Since the full Laplace--Beltrami operator is self-adjoint in the ambient Hilbert space, its adiabatic projection onto the degenerate subspace remains self-adjoint to leading order in the thin-waveguide
expansion. The projected longitudinal dynamics therefore consistently reorganizes into the gauge-covariant kinetic operator
\begin{align}\label{AppBDs2}
D_s^2&=(P\partial_s P)^2=(\partial_s+\Gamma_s)^2\notag\\&=\partial_s^2+2\Gamma_s\partial_s+(\partial_s\Gamma_s)+\Gamma_s^2,
\end{align}
where $P=\sum_c|\chi_c\rangle\langle\chi_c|$ denotes the adiabatic projector onto $\mathcal{H}_{\perp}^{(1)}$, and $D_s=P\partial_s P=\partial_s+\Gamma_s$ is the associated covariant derivative. Equivalently,
\begin{equation}
-\hbar^2 D_s^2=(-\mi\hbar\partial_s-\mathcal{A}_s)^2.
\end{equation}
In the adiabatic limit, the projected longitudinal dynamics is governed by the covariant operator $D_s^2$ up to nonadiabatic corrections discussed below.

We now examine the cross term $\Delta_{\text{cross}}$, which contributes to the emergence of the covariant kinetic structure in the curvilinear Frenet coordinates. Under the thin-waveguide condition $\kappa\ell_\perp\ll1$, we set $f\simeq1$ inside the projected matrix elements and retain the leading-order terms. The cross operator then reduces to
\begin{align}
\Delta_{\text{cross}}
\simeq
&\,
\partial_s(\tau v\partial_u)
-\partial_s(\tau u\partial_v)
+\partial_u(\tau v\partial_s)
-\partial_v(\tau u\partial_s).
\end{align}
Note that the derivative $\partial_s$ in terms like $\partial_s(\tau v \partial_u)$ acts on everything to its right, contributing to both the $\tau'(\dots)$ and $\tau(\dots)\partial_s$ structures, which is essential for assembling the covariant derivative.
Acting on the wavefunction $\Psi=\sum_b\psi_b(s)\chi_b(u,v;s)$, the first term gives
\begin{align}
\partial_s(\tau v\partial_u\Psi)
&=\tau' v\sum_b\psi_b\partial_u\chi_b\notag\\&+
\tau v\sum_b\left[(\partial_s\psi_b)\partial_u\chi_b
+
\psi_b\partial_s\partial_u\chi_b\right].
\end{align}
Projecting onto $\chi_a$ yields
\begin{align}
&\langle\chi_a|\partial_s(\tau v\partial_u)\Psi\rangle_{\perp}=
\tau'\langle\chi_a|v\partial_u|\chi_b\rangle_{\perp}\psi_b
\notag\\
+&\tau\langle\chi_a|v\partial_u|\chi_b\rangle_{\perp}\partial_s\psi_b
+\tau\langle\chi_a|v\partial_u\partial_s|\chi_b\rangle_{\perp}\psi_b.
\end{align}
Using the explicit oscillator states, the nonvanishing matrix elements are
\begin{align}
\langle\chi_2|v\partial_u|\chi_1\rangle_{\perp}
=&C^2\int\dif u\dif v\,v^2(1-\alpha u^2)\me^{-\alpha(u^2+v^2)}\notag\\
=&C^2\Bigl(\frac{\pi}{2\alpha^2}-\frac{\pi}{4\alpha^2}\Bigr)=\frac{1}{2},
\end{align}
and similarly
\begin{align}
\langle\chi_1|v\partial_u|\chi_2\rangle_{\perp}
=
-\langle\chi_1|u\partial_v|\chi_2\rangle_{\perp}=
\langle\chi_2|u\partial_v|\chi_1\rangle_{\perp}=-\frac{1}{2},
\end{align}
while all diagonal matrix elements vanish by parity:
\begin{align}
&\langle\chi_1|v\partial_u|\chi_1\rangle_{\perp}=
\langle\chi_2|v\partial_u|\chi_2\rangle_{\perp}\notag\\=&
\langle\chi_1|u\partial_v|\chi_1\rangle_{\perp}=
\langle\chi_2|u\partial_v|\chi_2\rangle_{\perp}=0.
\end{align}
Thus the mixed-derivative operators generate purely off-diagonal matrix elements proportional to the same $\sigma_y$ structure as the connection $\Gamma_s$.

The remaining terms in $\Delta_{\text{cross}}$, namely those proportional to $\psi_b$ without longitudinal derivatives, involve matrix elements of the form $\langle\chi_a|v\partial_u\partial_s|\chi_b\rangle_{\perp}$ and $\langle\chi_a|u\partial_v\partial_s|\chi_b\rangle_{\perp}$,
together with additional contributions generated by integration by parts. Using the Frenet--Serret evolution of the co-rotating basis, these terms preserve the same off-diagonal $\sigma_y$ structure as the Berry connection $\Gamma_s$. Combined with the longitudinal projection $P\partial_s^2P$, they reorganize consistently into the gauge-covariant kinetic operator
\begin{equation}
D_s^2=(\partial_s+\Gamma_s)^2
\end{equation}
to leading order in the combined thin-waveguide and adiabatic
expansion. Thus the mixed derivatives do not introduce an independent geometric interaction, but ensure the consistency and Hermiticity of the projected Laplace--Beltrami operator in the curvilinear Frenet frame.

Consequently, within this leading-order approximation, the projected longitudinal kinetic operator is consistently represented by the covariant form
\begin{align}\label{AppBcov}
&\left\langle\chi_a\left|-\frac{\hbar^2}{2m}(\Delta_{\parallel}+\Delta_{\text{cross}})\right|\chi_b\right\rangle_{\perp}
\notag\\
=&\frac{1}{2m}\Bigl[-\hbar^2\delta_{ab}\partial_s^2+2\mi\hbar\mathcal{A}_{ab}\partial_s+\mi\hbar(\partial_s\mathcal{A}_s)_{ab}+(\mathcal{A}_s^2)_{ab}\Bigr]\notag\\
=&\frac{1}{2m}(-\mi\hbar\partial_s-\mathcal{A}_s)^2.
\end{align}

Finally, we address the nonadiabatic correction $Q_{ab}$ omitted in Eq.~\eqref{AppBDs2}. From the identity
\begin{equation}
\langle\chi_a|\partial_s^2\chi_b\rangle_{\perp}=(\partial_s\Gamma_s)_{ab}+(\Gamma_s^2)_{ab}-Q_{ab},
\end{equation}
with $Q_{ab}=\langle\partial_s\chi_a|(1-P)|\partial_s\chi_b\rangle_{\perp}$.
The term $Q_{ab}$ represents virtual transitions to states outside $\mathcal{H}_{\perp}^{(1)}$. In the adiabatic thin-waveguide regime, the basis states $\chi_a$ are chosen to co-rotate with the Frenet frame, so $\partial_s\chi_a$ has only a small component orthogonal to the subspace, suppressed by factors such as $\kappa\ell_\perp$ and $\tau\ell_\perp$. Consequently, $Q_{ab}$ is of higher order in the small parameters and is neglected in the leading-order effective Hamiltonian.

Including the curvature-induced geometric potential $-\frac{\hbar^2}{8m}\kappa^2(s)$ (which is proportional to the identity in the degenerate subspace, as derived in Appendix~\ref{scalar}), the effective Hamiltonian is
\begin{equation}
H_{\text{eff}}=\frac{1}{2m}(-\mi\hbar\partial_s-\mathcal{A}_s)^2-\frac{\hbar^2}{8m}\kappa^2(s).
\end{equation}
This establishes the emergence of the torsion-induced gauge structure in the leading-order adiabatic dynamics of a thin waveguide with finite transverse width.

\section{WKB reduction of the covariant Euler--Bernoulli operator}
\label{WKB}

Here we justify the replacement $D_s^4 \to (k\mp\tau)^4$ used in the derivation of the acoustic dispersion relation \eqref{dispersion_acoustic}. The starting point is the minimal covariant effective equation \eqref{EB_covariant},
\begin{equation}
\rho A\,\partial_t^2 \boldsymbol{\psi} + EI\,D_s^4 \boldsymbol{\psi} = 0,
\quad
D_s = \partial_s + \Gamma_s,
\label{EB_start}
\end{equation}
where $\Gamma_s = -\tau(s)\mi\sigma_y$.
We seek solutions in the WKB (adiabatic) form $\boldsymbol{\psi}(s,t) = \boldsymbol{\phi}(s)\,\me^{\mi(k s-\omega t)}$,
where the envelope $\boldsymbol{\phi}(s)$ is assumed to vary slowly compared with the carrier wavelength $\lambda=2\pi/k$, and the torsion $\tau(s)$ is assumed to vary slowly on the same scale. The small adiabatic parameter is
\begin{equation}
\epsilon \equiv \frac{1}{k L_{\tau}} \ll 1,
\qquad
L_{\tau} \equiv \left|\frac{\tau}{\tau'}\right|,
\label{adiabatic_param}
\end{equation}
with $L_{\tau}$ the characteristic length scale of torsion variation.

Acting with the covariant derivative on the ansatz gives
\begin{align}
D_s \boldsymbol{\psi}
&= \bigl(\partial_s \boldsymbol{\phi} + \mi k \boldsymbol{\phi} + \Gamma_s \boldsymbol{\phi}\bigr)\me^{\mi(k s-\omega t)}
\notag\\&= \bigl(\mi k + \Gamma_s + \partial_s\bigr)\boldsymbol{\phi}\,\me^{\mi(k s-\omega t)} .
\label{D_action}
\end{align}
Because $\boldsymbol{\phi}(s)$ is slowly varying, $\partial_s\boldsymbol{\phi} = O(\epsilon k\boldsymbol{\phi})$. The WKB requirement is not that $\tau/k$ be small, but that $\tau(s)$ and the envelope vary slowly on the scale of the local wavelength. The principal symbol of $D_s^4$ is therefore obtained by replacing $\partial_s+\Gamma_s$ with $\mi k+\Gamma_s(s)$ locally. To leading order in $\epsilon$, the operator acting on the envelope is
\begin{equation}
D_s \boldsymbol{\psi}
= \bigl[\mi k + \Gamma_s + O(\epsilon k)\bigr]\boldsymbol{\phi}\,\me^{\mi(k s-\omega t)} ,
\label{D_leading}
\end{equation}
where $O(\epsilon k)$ collects envelope-gradient and torsion-gradient contributions that are subleading in the adiabatic expansion.

We now diagonalise $\Gamma_s$ in the circular-polarization basis $\boldsymbol{e}_{\pm}(s) = \frac{\boldsymbol{N}(s)\pm\mi\boldsymbol{B}(s)}{\sqrt{2}}$,
which are eigenvectors of $\sigma_y$ with eigenvalues $\pm 1$. Expanding the envelope as $\boldsymbol{\phi}(s)=\sum_{\alpha=\pm}\phi_{\alpha}(s)\boldsymbol{e}_{\alpha}(s)$, the action of $\Gamma_s$ on each component is
\begin{equation}
\Gamma_s \boldsymbol{e}_{\pm}
= -\tau(s)\mi\sigma_y \boldsymbol{e}_{\pm}
= \mp\mi\tau(s)\boldsymbol{e}_{\pm},
\label{Gamma_eigen}
\end{equation}
where the upper sign corresponds to $\boldsymbol{e}_{+}$ and the lower sign to $\boldsymbol{e}_{-}$. Consequently, to leading order,
\begin{equation}
D_s \phi_{\pm}\boldsymbol{e}_{\pm}
= \bigl[\mi(k\mp\tau) + O(\epsilon k)\bigr]\phi_{\pm}\boldsymbol{e}_{\pm}\,\me^{\mi(k s-\omega t)} .
\label{D_component}
\end{equation}

Iterating the derivative four times gives
\begin{equation}
D_s^4 \phi_{\pm}\boldsymbol{e}_{\pm}
= \bigl[\mi(k\mp\tau)\bigr]^4 \phi_{\pm}\boldsymbol{e}_{\pm}\,\me^{\mi(k s-\omega t)} + O(\epsilon k^4) .
\label{D4_component}
\end{equation}
Because $[\mi(k\mp\tau)]^4 = (k\mp\tau)^4$, the leading-order projected equation for each circular component becomes
\begin{equation}
-\rho A\,\omega^2 \phi_{\pm} + EI\,(k\mp\tau)^4 \phi_{\pm} = 0,
\label{dispersion_proj}
\end{equation}
which yields the dispersion relation \eqref{dispersion_acoustic},
\begin{equation}
\omega_{\pm}(k) = \sqrt{\frac{EI}{\rho A}}\,(k\mp\tau)^2 .
\label{dispersion_final}
\end{equation}

The corrections neglected in this leading-order treatment arise from two sources:
\begin{enumerate}
\item[(i)] \textbf{Envelope variation.} The term $\partial_s\boldsymbol{\phi}$ in Eq.~\eqref{D_action} contributes to $D_s^4$ at relative order $O(\epsilon)$. In the WKB hierarchy, this is subleading compared with the carrier term $\mi k$.
\item[(ii)] \textbf{Torsion gradient.} The derivative $\partial_s$ also acts on $\Gamma_s$ through $\partial_s(\tau\sigma_y)=(\partial_s\tau)\sigma_y$. The resulting terms in $D_s^4$ are proportional to $\tau'$, $\tau''$, etc., and are of relative order $O(\epsilon^2)$ or smaller under the adiabatic condition \eqref{adiabatic_param}.
\end{enumerate}
Both corrections modify the dispersion at subleading order in the adiabatic expansion but do not alter the principal momentum shift $k\to k\mp\tau$ generated by the geometric connection. Thus, Eq.~\eqref{dispersion_final} is the principal WKB symbol of the covariant Euler--Bernoulli operator \eqref{EB_start}.

\bibliography{Review1}

\end{document}